\documentclass[journal]{IEEEtran}
\usepackage{graphicx}
\usepackage{amstext}
\usepackage{algorithm}
\usepackage{algorithmic}
\usepackage{mathrsfs}
\usepackage{amssymb}
\usepackage{amsmath}
\usepackage{bm}
\allowdisplaybreaks[4]
\usepackage{epstopdf}
\usepackage{multicol}
\usepackage{stfloats}
\usepackage{url}
\usepackage{color}
\usepackage{enumerate}
\usepackage{graphics}
\usepackage{multirow}
\usepackage{graphicx}
\usepackage{mathtools}
\usepackage[utf8]{inputenc}
\usepackage{multirow}
\usepackage{amsmath}
\usepackage{relsize}
\usepackage[usenames,dvipsnames]{xcolor}
\usepackage{tcolorbox}
\usepackage{tabularx}
\usepackage{array}
\usepackage{colortbl}
\tcbuselibrary{skins}
\usepackage{cite}
\usepackage{algorithmic}
\usepackage{amssymb}
\usepackage{caption}
\usepackage{lipsum}
\usepackage{xcolor}
\usepackage{float}
\usepackage{amsthm}
\usepackage{graphicx}
\usepackage[tableposition=top]{caption}
\usepackage{breqn}
\usepackage[font=small,skip=0pt]{caption}
\usepackage{setspace}
\usepackage[shortlabels]{enumitem}
\usepackage{float}
\usepackage{caption}
\usepackage{romannum}
\usepackage{makecell}
\usepackage{longtable,booktabs}
\usepackage{tikz}
\usepackage{newunicodechar}
\usepackage{color,soul}
\usepackage{array}
\usepackage{lipsum}
\usepackage{romannum}
\usepackage{lscape} 
\usepackage{longtable}
\usepackage{graphicx}
\begin{document}
\title{Federated Learning for Privacy Preservation in Smart Healthcare Systems: A Comprehensive Survey  }

\author{Mansoor Ali,~\IEEEmembership{Member,~IEEE,}
        Faisal Naeem,~\IEEEmembership{Member,~IEEE,}
		Muhammad Tariq,~\IEEEmembership{Senior Member,~IEEE,}
        and Geroges Kaddoum,~\IEEEmembership{Senior Member,~IEEE,}
        \thanks
		{Mansoor Ali, Faisal Naeem and Georges Kaddoum iare with the Electrical Engineering Department, ETS, University of Quebec, Montreal, Canada. (E-mail: mansoor.ali.1@etsmtl.net, faisal.naeem.1@etsmtl.net, and  georges.kaddoum@etsmtl.ca) 
		}
		\thanks{Muhammad Tariq is with the Department of Electrical and Computer Engineering, Princeton University, NJ, USA, 08544. E-mail: (email:
			mtariq@princeton.edu)}
		 \thanks {Corresponding Author: Mansoor Ali (mansoor.ali.1@etsmtl.net)}
\thanks{}
\thanks{}
\thanks{}
\thanks{}%
}

\maketitle

\begin{abstract}
Recent advances in electronic devices and communication infrastructure have revolutionized the traditional healthcare system into a smart healthcare system by using internet of medical things (IoMT) devices. However, due to the centralized training approach of artificial intelligence (AI), mobile and wearable IoMT devices raise privacy issues concerning the information communicated between hospitals and end-users. The information conveyed by the IoMT devices is highly confidential and can be exposed to adversaries. In this regard, federated learning (FL), a distributive AI paradigm, has opened up new opportunities for privacy-preservation in IoMT without accessing the confidential data of the participants. Further, FL provides privacy to end-users as only gradients are shared during training. For these specific properties of FL, in this paper, we present privacy-related issues in IoMT. Afterward, we present the role of FL in IoMT networks for privacy preservation and introduce some advanced FL architectures incorporating deep reinforcement learning (DRL), digital twin, and generative adversarial networks (GANs) for detecting privacy threats. Moreover, we present some practical opportunities of FL in IoMT. In the end, we conclude this survey by providing open research challenges for FL that can be used in future smart healthcare systems. 

\end{abstract}


\IEEEpeerreviewmaketitle

\section{Introduction}
Internet of medical things (IoMT) has transformed healthcare systems into personalized, user-centric, precise, and ubiquitous services by providing improved healthcare management, supervision and procedure by improving the quality of life and human well-being \cite{tai2021trustworthy}. The smart healthcare system consists of IoMT devices widely used to monitor medical traffic continuously. The healthcare traffic is forwarded to artificial intelligence (AI) enabled framework to realize a plethora of emerging intelligent healthcare systems, such as disease prediction and remote health monitoring \cite{mansour2021artificial}. The future healthcare systems will be based on applications such as holographic communication, telesurgery, Hospital-to-Home (H2H), and Quality of Life (QoL) services. In particular, telesurgery and holographic communication will have stringent and real-time performance requirements. Due to low data rates, existing wireless architectures such as 5G cannot support intelligent healthcare applications. The 6G is expected to play a vital role in completely revolutionizing the existing healthcare system and overcoming the communication barriers of the existing wireless architecture.
	
	Traditionally, healthcare traffic management techniques rely on a centralized AI framework that is located in the data center or cloud for health data analytics and learning. Given the growth of IoMT devices and large volumes of health traffic in modern smart health care systems, the centralized architecture results in scalability issues \cite{naeem2020sdn}. Furthermore, the reliance on centralized servers for data learning makes them vulnerable and exposed to various types of security threats and thus, poses a significant threat to patient's safety and privacy \cite{yuan2020federated}. For example, adversaries and intruders can hack the IoMT devices and modify the patient has stored data resulting in a threat to the patient's life. Moreover, the centralized AI framework for future adaptive healthcare systems may not be practical because health traffic will be distributed over heterogeneous and large-scale healthcare systems. As a result, for the successful deployment of intelligent healthcare systems, a distributed AI-based framework is needed for enabling privacy-preserving and scalable healthcare applications.  

    In this context, the emerging concept of federated learning (FL) has shown promising solutions for providing privacy protection in smart-healthcare networks \cite{sheller2020federated,nguyen2021federated}. FL is a type of distributed AI that enables the models' distributed training by averaging the local model updates from multiple IoMT networks without accessing the local data. As a result, potential risks of disclosing user preferences and sensitive patient information can be mitigated by using the FL approach. Moreover, the FL approach improves the training performance of the health by collecting large datasets and computation resources from local IoMT devices, which might not be possible in the case of using the centralized AI approach \cite{nguyen2021federated}.

\begin{table*}[t]

\caption{Comparison of existing survey in FL in smart healthcare}
 \begin{tabular}{|p{1.5cm}|p{7cm}|p{8cm}|} 
 \hline
    \rowcolor{gray}

Reference & Area of focus  & Our contributions                     \\ \hline
  \cite{khan2021federated}& Outlines the applications of FL in security and privacy in IoT networks. & \multirow{5}{*}{\begin{tabular}[c]{@{}l@{}} \\ \\ This survey provides a holistic taxonomies and a \\ comprehensive survey on the applications of FL \\ in IoMT network from the perspective of privacy \\ preservation. \end{tabular}}  \\ \cline{1-2}
\cite{xu2021federated}  & In the survey paper, FL is explored within the healthcare framework, but privacy and security concerns are not addressed.&                                       \\ \cline{1-2}
\cite{pham2021fusion}   & Focuses on the role of FL in Industrial Internet of Things (IIoT) and discussion of FL in healthcare is very limited.          &                                       \\ \cline{1-2}
 \cite{rieke2020future} &  A discussion of the requirements and technical issues of FL in digital health.      &                                       \\ \cline{1-2}
 
\cite{nguyen2021federated}& Summarizes the requirements and role of FL in healthcare. &     \\ \hline
\end{tabular}

\label{table-intro}

\end{table*}

\subsection{Comparison and Our Contributions}
Driven by recent advances in FL, many researchers have surveyed the area of FL in healthcare systems. The work in \cite{khan2021federated} conducted a detailed comparative analysis of applications of FL related to security in IoT networks. The paper in \cite{xu2021federated} performed a survey of FL in the area of healthcare networks. However, the security perspective is not covered. Meanwhile, the role of FL in technical implementation in digital health is explored in detail in \cite{rieke2020future}.
Moreover, a recent paper \cite{nguyen2021federated} discussed the role of FL in intelligent healthcare and covered some aspects of privacy. However, the recent advances of the FL in privacy-preserving schemes, including the recent development of AI techniques such as reinforcement learning (RL), digital twin (DT), and generative adversarial network (GAN), have not been explored to the best of our knowledge. The overall comparison of this work with the recent surveys is summarized in Table 1.

The remainder of the survey is organized as follows. Section II presents the security and privacy issues in IoMT; Section II describes FL's motivation and architecture in IoMT networks. Section IV introduces some advanced architectures of FL from the privacy perspective. In Section V, we have described the application of FL followed by Section VI, where research directions are given. Finally, the survey has concluded Section VII.

\section{Security and privacy of IoMT}
Many researchers in the last decade have addressed the literature concerning IoT privacy and security. It has been observed that addressing the security issues was primarily focused \cite{1}; however, more attention needs to be paid to the protection of end-user privacy in modern electronic health care systems \cite{2,3}. In the modern era, the health care system is transformed into a new domain by incorporating advanced digital technologies such as IoT, high computing devices to store and process data, personal health records, and more. Through these advanced countermeasures, it has been observed that IoMT can significantly improve the efficiency of the healthcare system by ensuring the safety of patients \cite{4,5}.

Despite all of these advantages, cybersecurity of the healthcare system is an important issue that needs to be tackled \cite{6}. Due to poor defense systems for information security, electronic healthcare systems are considered an easy target for attackers. An adversary can hack into the system, making it paralyzed using ransomware, getting patient information from hospitals and selling it, and blackmailing patients into releasing their personal data. Furthermore, it has been observed that attack-able vulnerabilities points are more prevalent in the electronic healthcare system. One such instance can be found in \cite{7}, where the attackers intercept the operation of digital insulin pumps. The security operator reported such an attack by analyzing the communication protocols. Thus, it is observed that before deploying any security and privacy algorithm, risk analysis of the system is the foremost priority. Henceforth, security and privacy issues pose a significant threat to modern healthcare systems and need to be addressed for the effective operation of IoMT systems. Moreover, the secure network must fulfill all the system requirements under constraints and must be adaptable to changes in the healthcare system to meet future needs.

\subsection{Privacy in IOMT}
In today's world securing the network is not sufficient for its effective operation. Protection of networks from a privacy perspective is also important, specifically for electronic healthcare systems. Here, in this section, we will provide a brief description of privacy threats and relative effective protection schemes.

\subsubsection{Private data}
In IoMT based applications, an abundant amount of data is being communicated among different operating systems connected to the network, raising privacy concerns. Therefore, various standard protocols such as ISO/IEC 27018 \cite{8}, and 29100 \cite{9} are designed to tackle these issues to some extent. The information type that is used to present the data regarding the specific individual is referred to as Personal Identifiable Information (PII) \cite{9}. Based on that, the user data can be classified into three categories such as sensitive personal data, general data, and statistical data, respectively. As its name implies, sensitive personal information requires the highest level of privacy, while general and statistical data require moderate protection since they are primarily used for surveys and statistical analysis.

The PII owner is the one to whom the information belongs, and they have complete authority over it. In contrast, a PII-associated processor is an organization to whom the person has given the rights to access their personal information and use them for various purposes. Nevertheless, the processor might share the information with a third party after getting consent from the PII owner for certain specific functionalities. If any violation involves using the personal information for any unauthorized purpose, the contracted processor and third party will be held liable.

\subsubsection{Prevention mechanisms}
Malicious events that compromise the privacy of PII could include tampering with the connection between smart devices in the home or access to private information via wearable health monitoring devices \cite{10,11}. Thus, private data must be protected before sharing and storing it. However, there is some protection mechanism that can be adopted to protect the data throughout its processing steps from collection to storage and analysis. 

The data that is collected for monitoring purposes from patients is abundant; hence for IoMT related devices, definite control steps must be deployed \cite{12}. For instance, the data collection that is requested by the healthcare application must be reduced to reduce computation costs. Moreover, storing all data must be minimized for an extended period and enforcing data to store for a short period. Furthermore, encouraging data processing on edge servers to reduce computation time and process more data while protecting user identity. For privacy concerns, it is more suitable to adopt anonymization on user data to hide PII personal information to secure data leakage to unintended users.

\subsubsection{Privacy of IoMT using identification and anonymity}
The main concern of any preservation techniques adopted for privacy is to secure the user identity and information. The adversary may get the user information by analyzing the data flow among different individuals. For this purpose, the users may choose to maintain their anonymity from the network operator for some applications and access services while preserving their privacy. The three types of secure access that users can adopt are: adopting login bases authentication of the user, using a pseudonym, and utilizing anonymity. The user is aware of the fact that the services provider might track the user identity if login-based privacy is used. Based on this fact, third parties might use the available user information. In such circumstances, to avoid data leakage, data encryption is highly recommended. 

The users can hide their critical data by utilizing pseudonyms, providing privacy protection of high levels for many applications. However, the adversary can infer information of an entity by using context-based knowledge algorithms. For instance, the hospital operators can identify whether the users are employees, patients, or other personnel related to the patient that put the hospital services request. Moreover, the personnel can be identified as staff operators if they use IoT-enabled hospital services almost daily. Furthermore, if the user is accessing the services from remote locations, then the adversary can easily deduce that the user is using the facilities from their home and will try to get the identity of the users through context-based identification algorithms. Henceforth, more prevention mechanisms are needed for location-based services (LBS), especially where the IoT network is involved \cite{13}.

In order to overcome the loopholes in pseudonyms, more robust mechanisms like cloaking areas and k-anonymity were adopted in \cite{14,15}. Cloaking areas imply that pseudonyms for various users are interchanged in a random manner when passing through a particular area. For instance, in IoT environments where smart cars are included, anonymization is usually performed in areas where more cars are gathered, like traffic signals or road crossings. However, it was observed that by using a context-aware knowledge algorithm, the user identity could be inferred \cite{16}. Hence, more robust protection methods like semantic obfuscation algorithms are adopted to counter these attacks \cite{17}. By adopting K-anonymity, the user identity can be blurred from the service provider by the intermediate entity. However, the intermediate entity must be carefully selected and should be trustworthy among the users. Furthermore, for privacy and security purposes, the different functional operations are mostly carried out on user devices; hence the resources of active participants are only used. Nevertheless, by adopting k-anonymity, the privacy protection levels can be controlled and quantified, which is the main advantage of the anonymity algorithm \cite{18}.

\section{FL and its perspective in IoMT}

The introduction of the machine learning (ML) algorithm is gaining more attraction because of its complex modeling capability from large datasets stored on a central server \cite{19}. In traditional ML, the data is mostly stored at a central location without considering any privacy prevention countermeasures and data transmission costs. For that purpose, privacy assurance can be achieved by using Blockchain, and authentication algorithms \cite{20,21}. Moreover, it has been observed that ML-based intrusion detection algorithms can also be used for the identification of adversary actions \cite{22,23}. Nowadays, ML-based privacy preservation is mostly used due to its efficiency \cite{24}, and the more preferable technique is federated learning (FL). Initially, Google used FL as a means to combat the privacy concerns with traditional ML, as FL works collaboratively on end devices \cite{25}. However, traditional FL has privacy issues as local model updation depends upon central model parameters. If the central global model is compromised, the overall efficiency of the whole FL framework suffers \cite{26, 27, 28}. Moreover, there are constraints associated with individual IoT devices, such as power and resources, limiting the FL operation. These constraints need to be optimized to ensure efficient operation of FL \cite{29}.


The applications of FL are found in various major fields such as in network optimization \cite{30}, Google advanced keyboard for prediction \cite{31}, in healthcare systems like COVID-19 detection \cite{32} and in intrusion detection \cite{33}. However, finding the appropriate testing model for privacy and security algorithms for IoMT is still an open research issue, and researchers are finding it difficult to either use a centralized or FL-based model to test \cite{34}. The main difference between centralized ML and FL is that 
in centralized ML the learning data is uploaded to the central server, where processing is performed and information is shared among different users. However, the main issue with that learning is privacy and management of the massive flux of data transmitted by IoMT devices. In contrast to centralized ML, in FL, the learning is performed in a distributed fashion on end devices data and model parameters are transferred to the central global model, thus ensuring the privacy of data shared and optimal traffic management in IoMT networks. Moreover, there are different architectures in which FL is performed, such as horizontal FL, vertical FL, and transferred FL. The detailed architectures and working of this sub-framework can be found in our previous work \cite{28}.

\begin{figure*}[t]
\centering
\includegraphics[height=4in,width=5.5in]{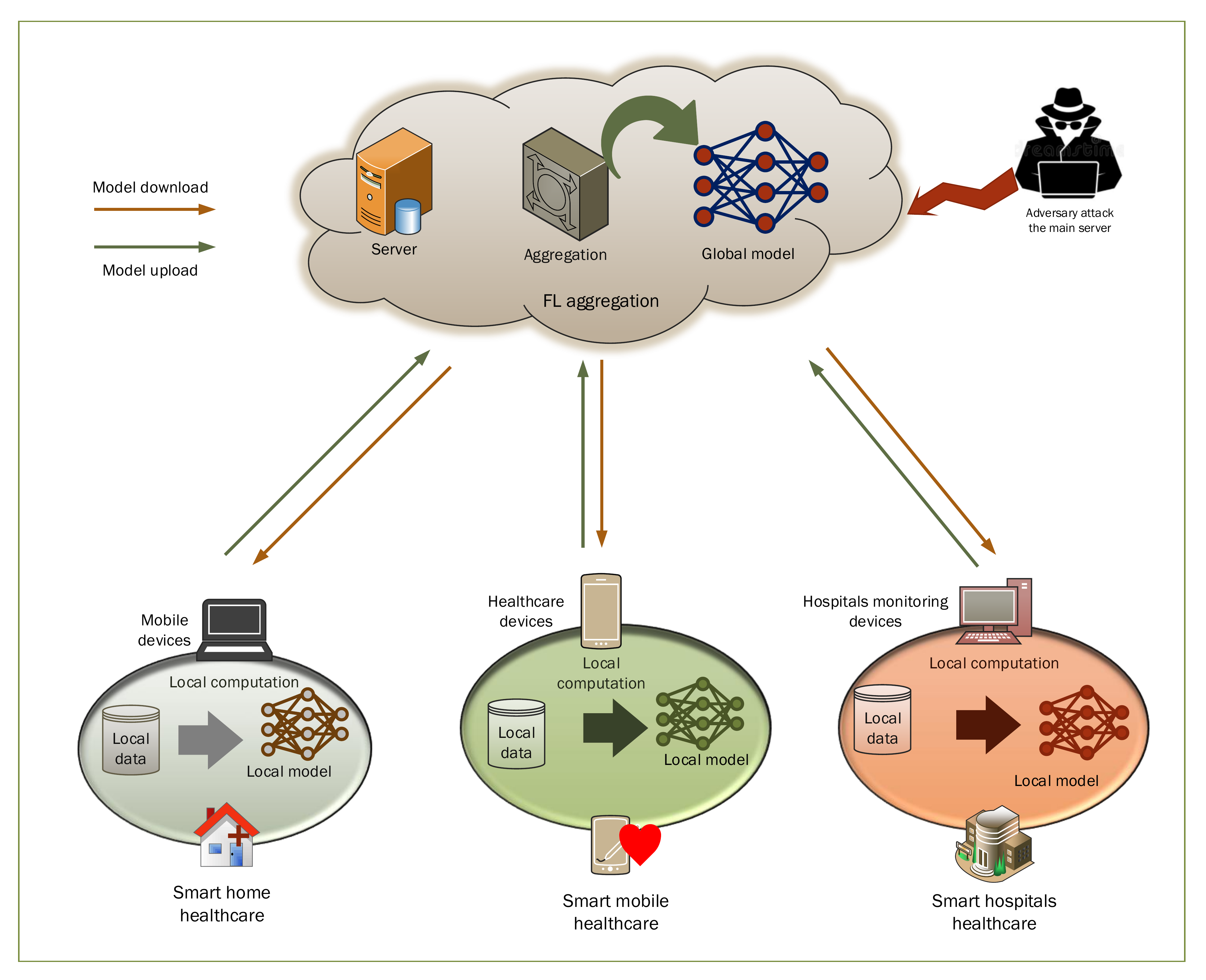}
\caption{FL based healthcare architecture}

\end{figure*}

\subsection{Framework for FL based healthcare system}

The basis architecture framework in which FL can be incorporated is shown in Fig. 1. The generalized FL-enabled healthcare system comprises certain steps. In the initial step, the central server selects different network parameters related to healthcare systems, such as deciding the task, whether medical image processing or some other human-related application and identifying the algorithm based on prediction or classification task. Moreover, the learning rates and different configurable parameters related to ML are selected. Furthermore, the central server decides the clients participating in the FL process. 

Once the central server decides the number of end nodes to participate in FL, then it shares the initial models among the nodes. The end nodes then train the model based on their local data, and the updated model is then shared back with a central model for aggregation. For instance, we can use a federated average model for aggregation, where the weights are assigned to local model parameters based on the data size availability \cite{35}. In the end, the new global model is computed and is shared back with the end nodes. The learning process is continuous in an iterative manner until the desired accuracy is achieved.

\subsection{Limitations in existing healthcare systems}

Some of the limitations, especially from a privacy perspective and traditional ML, are already discussed in Section. II. In this subsection, we will be discussing some other limitations that hinder the implementation of IoMT in real-world applications. For instance, in traditional ML approaches for healthcare data analysis, the data is stored on a central cloud and is vulnerable to adversary attacks. Moreover, a third-party platform is used for data processing and storage, which can access the data and may infer private information \cite{36}.

In a real-world healthcare system, the data gathered from a single medical lab is not enough, and the ML model based on that dataset will not give promising results \cite{37}. Thus, to tackle that, in most cases, manual analysis of the data is performed, which is an inefficient way of data processing. For this purpose, one possible way is by sharing data among different medical institutes, but given the fact of privacy, it is not easy to get the data for training the network. Moreover, insufficient data makes it impossible to train an accurate model since the input data is imbalanced and lacks enough feature information. 

\subsection{Role of FL in IoMT}

The new advanced features of collaborative learning in FL enable its application in various sectors, especially in smart IoMT. By adopting FL algorithms in IoMT, the local model parameters are communicated while the host data remains within the local nodes. This increases privacy and reduces information leakage scenarios. Moreover, the training of networks on diverse data increases the generalization capability of FL. Furthermore, the communication cost is reduced by uploading only gradients rather than larger datasets \cite{38}. These actions adopted by FL enable efficient bandwidth utilization and avoid network congestion in massive IoMT networks. Although FL opens a new domain for effective utilization of IoMT networks in the actual domain. There are some privacy issues associated with FL. For example, if an adversary gets access to the central server, then user information can be extracted, as shown in Fig. 1. Moreover, the detailed analysis of privacy issues in FL based IoMT network and proposed prevention algorithms are discussed in the subsequent sections.

\section{Featured FL design for IoMT}
In this section, we will discuss advanced FL architectures for IoMT networks from different perspectives that will be summarized in Table.2. 

\subsection{Privacy enabled FL}
The adoption of FL can resolve many privacy-concentrated issues in IoMT, but FL does have some privacy problems. For instance, the global model updates itself based on local model data from IoMT devices. The adversary can attack and get the user information by using construction attacks \cite{39}. Furthermore, by using an inference attack by the malicious user, the adversary can get what kind of information is being shared. This might include blood samples, type of disease and other relevant data \cite{40}. Moreover, the detailed literature regarding intrusion attacks can be found in \cite{agrawal2021federated,tahir2021experience}. Thus before developing any privacy mechanism, , the nature of IoMT attacks must be considered. 
\subsubsection{Information leakage}
The issue of information leakage by adopting a collaborative machine learning algorithm was addressed by authors in \cite{41}. In \cite{41}, the authors generated the attacks patterns by using generative adversary networks (GANs) against which the protection mechanism is designed. To tackle this attack, the robust FL mechanism called EaSTFLy was proposed in \cite{42} and was verified against IoT networks. In the EaSTFLy algorithm, two different privacy protocols, such as Paillier homomorphic encryption (PHE) and Shamir’s threshold secret sharing (TSS), are used to prevent the information leakage issue in FL algorithms.

\subsubsection{Poisoning attack}
In this kind of attack, the adversary tempered the local model updates parameters to reduce the aggregator accuracy. The authors in \cite{43} go one step ahead and introduce two types of poisoning attack, namely, data and model poisoning. In a data poisoning attack, the training data is tempered, while model parameters are attacked in model poisoning. To tackle such type of attack, GANs based prevention mechanism was introduced in \cite{44}. The proposed algorithm performs data auditing generated by GANS and performs comparative analysis to detect the malicious user.

\subsubsection{Byzantine attack}
The adversary or malicious node participation in FL shares the fake model parameters with neighbor nodes in this attack. Moreover, the attacker may reduce the convergence time and accuracy of the model by sharing false data. For this reason, Blockchain incorporated FL was presented in \cite{45} for prevention against Byzantine type of attacks. Moreover, in \cite{46} digital twin enabled FL was proposed, where FL even incorporated the malicious data into consideration to successfully prevent the privacy threats.

\subsubsection{Privacy data leakage attack}
In distributed FL, the model updates communicated with central servers by IoMT devices keep leaking some training data information. By initiating the differential attack, the adversary can find out whether the end devices are dedicated to a particular task or not. If so, the malicious user will get that information and use it for other purposes. The privacy prevention algorithm, composed of reinforcement learning, blockchain, and differential privacy, was presented in \cite{47} in order to combat these attack scenarios. The privacy prevention algorithm is deployed in a central aggregator to prevent privacy leakage attacks.

\subsubsection{Inference based attack}
This attack is mostly based on data mining. During this attack, the adversary implements data mining techniques to analyze the data and get some useful information from the data. The authors in \cite{48} adopted a privacy-preserving FL scheme to prevent such attacks, which ensured the security of the data of end devices both before and after training. Going one step further, differential privacy and homomorphic encryption-based federated gradient boost algorithm are proposed in \cite{49} to handle these inference-based attacks. Moreover, deep neural network (DNN) was adopted to detect cyber-attacks in IoMT \cite{rm2020effective}.

The major known technique that is mostly used to enhance privacy preservation in many areas is based on differential privacy. This property to prevent privacy leakage motivated many researchers to initiate working on differential privacy-based FL systems for IoMT and many other major areas. One of its applications can be found in \cite{50}, where the authors proposed a mechanism to deliberately add noise to the IoMT devices dataset to protect user private information. Moreover, a multi-dimensional incentive mechanism is designed based on cost optimization. For that purpose, three types of cost parameters are considered: computation, network communication, and privacy level cost, respectively. Through experimental analysis, it was concluded that multi-dimensional incentive mechanisms could give high accuracy by reducing training loss compared to the vanilla FL algorithm. Similarly, the work where differential privacy is adopted for federated IoMT applications can be found in \cite{51}. In \cite{51}, the authors proposed a methodology where stochastic gradient descent based differential privacy is used for distributed health data. To further secure private information, homomorphic encryption was adopted for aggregation. From a privacy preservation perspective, another application of differential privacy-based FL is investigated in \cite{52}. Based on the data, two parameters are used, including the effect of drugs prescribed and mortality rate and their probability of occurrence. It has been observed from the analysis of the results that by increasing privacy, the accuracy of the training algorithm is reduced. Hence, more research is needed to devise an FL algorithm that successfully increases the privacy and accuracy of the model on available data.

\subsection{Incentive enable FL for IoMT}
In vanilla FL algorithms, all the IoMT devices must participate in the training process and share their aggregation model, but in most cases, this does not occur. Due to the limited computation capability of IoMT devices and the privacy and trustworthiness of third-party platforms, IoMT devices do not show any willingness to contribute to FL approaches. To involve all the devices, incentives-based FL techniques are proposed. In a broad sense, the incentive mechanism for FL can be classified into three categories: incentive allocation based on end-device data, device participation, and resources available at end devices \cite{53}.

For designing incentive mechanisms for mobile network applications, game theory-based approaches are expected to be a good tool. Motivated by the good features of game-theoretic approaches different incentive mechanisms are proposed for FL in IoMT systems. For instance, the Stackelberg game was adopted to model the dynamic variation in end devices computation resources and stable connection between IoMT devices and central aggregator in hospital \cite{54}. The drawback with this algorithm is that one needs complete knowledge of FL devices and networks to deploy it fully. To tackle that drawback, a deep reinforcement learning (DRL) based algorithm is presented in \cite{55}, where DRL is responsible for allocation awards at the aggregator and selecting devices to suggest the learning model for their data. Furthermore, motivating different hospitals and companies to build the global model on the third-party server by contributing their resources to promote social welfare is an open problem. The authors in \cite{56} find the solution to this problem by arising two questions: how hospitals allocate resources to end devices and how other agencies participating in this program will compensate for the resources used by other hospitals. In this context, the problem is treated as non-convex optimization, and distributed techniques are investigated to solve such problems. From an effective incentive mechanism perspective, it was observed in \cite{57} that by using the Shapley value algorithm, the cost related to computing resources increases as the number of end devices and data used for training by them increases.

\subsection{FL enabled digital twin for IoMT}
It is now possible to find the application of digital twins (DT) in many fields, and one such application is discussed in \cite{58} in the area of IoMT. DT is said to be the digital replica of physical process and there is continuous data and information sharing between both entities. From the IoMT perspective, through DT virtual environment containing patients can be created and the doctors can test their prescribed medication in a virtual environment before using them on real-world patients. Furthermore, the incorporation of DT in IoMT enables secure remote patient monitoring (RPM). The long short term memory (LSTM) based anomaly detection was built-in \cite{58} and FL based DT was used for preventing adversaries from getting private information.
\begin{table*}[!t]\normalsize
\centering
\caption{Advanced FL architecture for IoMT network for privacy preservation}
\begin{tabular}{|p{0.9in}|p{0.4in}|p{0.7in}|p{0.6in}|p{0.7in}|p{1.2in}|p{1.2in}|}
\hline
                 \textbf{Theme} & \textbf{Ref}  & \textbf{FL type}  & \textbf{FL client nodes}  & \textbf{Aggregator server type}  &\textbf{Contribution}  & \textbf{Limitations}  \\ \hline
& \cite{50} & Horizontal FL (HFL)  & Smart IoMT nodes  & Cloud based aggregation & Multi dimensional cost optimization using incentive mechanism for FL privacy.  & Numerical methods are needed for performance evaluation.  \\ \cline{2-7} 
                  Privacy enabled FL&\cite{51}  & HFL  &Corporate hospitals  &Cloud based aggregation  &Privacy preservation and accuracy is gained using differential privacy and DNN.  &More robust and lightweight encryption algorithm is needed.   \\ \cline{2-7} 
                  & \cite{52} & HFL  & Smart mobile devices  & FL based servers  & A comparative analysis between accuracy and  privacy.  & FL algorithm has convergence issues and is not addressed.   \\ \hline
&\cite{54}  &HFL  & Smart IoMT nodes &FL based server  &Stackelberg game was adopted
to model the dynamic variation in end devices computation
resources and stable connection between IoMT devices.  & Numerical methods are adopted for performance evaluation. \\ \cline{2-7} 
                 Incentive enable FL for IoMT &\cite{56}  &HFL  &Corporate hospitals  & Data center  &A strategic game is played for resource allocation based on incentive mechanism.  & The models works on assumption made on the bases of independent and identically distributed (IID) data.  \\ \cline{2-7} 
                  &\cite{57}  &HFL  &Smart wearable devices  &FL based server  &The contribution of local model for training global model was evaluated based in DRL.  & An incentive based reward is needed to encourage the nodes for participation in FL.  \\ \hline
                  FL enabled digital twin for IoMT&\cite{58}& Hierarchical FL  & Smart healthcare devices & Centralized server  &DT was adopted for anomaly detection in main aggreagtor server.  &More robust DT based model is needed to identify unseen threats for privacy preservation.   \\ \hline
\end{tabular}
\end{table*}

\section{Application of FL in smart healthcare IoMT systems}
We have discussed privacy issues in IoMT and some of the prevention techniques based on FL in IoMT. In this section, we will explore some applications of FL in IoMT.

\subsection{Electronic health record management using FL}
The digital information that is stored in EHR, in most cases ML algorithms are used on these records for health assessment and diagnosis. However, traditional ML techniques have privacy concerns, which could lead to sensitive data being leaked during analysis. For security purposes, the patient name is removed from health records, but this solution is not sufficient as hospitals and agencies work in collaboration to find the solution of new diseases, and they commonly share a single central server where patient data is stored.

In such circumstances, FL-based techniques provide a more reliable intelligent solution to EHR management while preserving the privacy of patient data, especially where multiple cooperation data is involved. One such example was presented in \cite{59} in which privacy and optimal resource usage-based protocol schemes were presented for FL to analyze EHR. To further enhance the privacy in local model parameters, perturbation of training data is performed to prevent memorization attacks. The main advantage of this technique is that even if the adversary can get information about perturbations in EHR, the original information will remain safe and secured.  

\subsection{FL in medical image processing}
Nowadays, ML is used in many healthcare applications, such as in medical image processing. However, due to privacy concerns, many institutions avoid sharing their private data for processing. In such a case scenario, FL-based algorithms are considered to be the most viable option, as the model is trained on multiple datasets without sharing their private information. The application of FL-based image processing is found in \cite{60}. In this paper, the authors try to develop a single image from different images of clients using FL at the global server. This process enables the building of diffusion coefficient-based images depending on different sources and classifications. Moreover, GAN is used to create raw image datasets by multiple hospitals and then share those raw imprints rather than actual images for privacy preservation. It has been observed using experimental analysis that the algorithm provides 97\% accuracy on cancer datasets and outperforms other non-FL-based schemes. 

The application of FL for brain imaging is presented in \cite{61}. The client has enough data related to a brain tumor as well as enough resources to train a deep neural networks (DNN) model and share updated model parameters for aggregation at the central server. However, there is a risk of leaking the model data during communication between the server and local clients. So, to counter that issue, differential privacy is adopted through which noise is added to local model updates, which ultimately reduces the risk of information leakage during parameter sharing. 

\subsection{Role of FL in COVID- 19}
There has been a worldwide pandemic caused by the spread of COVID-19, which is considered a major health threat \cite{62}. To reduce the rapid spread of the disease, many ML-based algorithms are proposed for the early detection of COVID-19. In most cases, DNN based algorithms called convolutional neural networks (CNNs) are used for COVID-19 detection using feature extraction from a patient chest X-ray \cite{63}. However, the collection and sharing of abundant data for designing an effective training model is a challenging task due to privacy concerns and the fact that most user may not share their medical records for analysis purposes \cite{64}. Given these facts, FL is considered to be the most appropriate candidate for detecting COVID-19 while preserving user information. While using FL, each hospital trains their model based on locally available X-ray images and only shares gradients with a global model for computation, hence ensuring privacy preservation \cite{65}.

Moreover, authors in \cite{66} adopted a dynamic FL algorithm, which uses a two-step process for COVID-19 diagnosis that is client participation and selection, respectively. During each iteration, the hospital decides whether to participate in training or not based on the model performance. Furthermore, the central server decides which client's model will be selected for aggregation based on updating time. In case if a global server does not receive the local model parameters at the appropriate time, then that model is excluded from global aggregation.

\section{Research direction}

In this section, some open research problems related to IoMT health systems are discussed.

\subsection{Communication network issues in FL for IoMT}
The communication network is a crucial factor to consider when deploying FL for applications since model updates are often transmitted through it. Indeed optimal resource scheduling is very important if one wants to design an effective training model. The case scenario is more severe in the case of a smart healthcare system where a large number of IoMT devices are present. These devices share their model parameters using the uplink channel and get the aggregator model through the downlink network. Hence, a more sophisticated resource allocation algorithm is needed while considering the highly dynamic characteristics of the wireless network.

\subsection{Universal FL architecture for IoMT}
The performance and importance of FL in IoMT are studied in detail by researchers in the literature. However, to properly evaluate the FL algorithm some universal standards and protocols need to be presented. For instance, to reduce the functionality of the central server, many blockchain algorithms are presented for IoMT, but no comparative analysis has been found. This is mainly due to the fact that each algorithm is deployed under a different model specification, such as dataset and application-specific  \cite{67}.

\subsection{Need of robust FL for diffused health dataset}
In real-world applications, the data available by each client is unique; that is, it might be in different forms like images, video, or text, and they will be showing different information like blood type and sugar level, etc. More probably, all the FL privacy preserved algorithms are designed in such a manner that the clients will have a dataset containing almost the same features \cite{51,52}. Given such circumstances, a more robust heterogeneous FL mechanism is needed where clients having different specific datasets take part in sharing their local model, and the global aggregator can incorporate this dynamic model data by using ensemble learning algorithms \cite{68}.

\subsection{FL for next generation IoMT networks}

Even though the full potential of a 5G network is yet to be recognized by the world, many professionals are working on the deployment and development of a 6G wireless network \cite{69}. The adoption of the 6G network can be used in many industrial application sectors ranging from Industry 5.0 to body area networks. In a 6G network, the data generation and collection will be abundant. In such scenarios, the performance parameters of FL caused by 5G/6G networks and their application to the healthcare system are considered as an open research problem.

\section{Conclusion}
Despite the many potential applications of smart healthcare systems, their deployment in the real world is limited due to the lack of privacy protection. For this specific purpose, FL is considered to be a viable option as FL operates in a collaborative fashion and provides privacy to user information. However, given the fact of the importance of FL, a limited comprehensive survey regarding FL application in IoMT is available. In this paper, we have conducted a detailed survey about FL applications for privacy preservation in IoMT networks. Firstly, we have described privacy issues in IoMT and performed an analysis of available privacy preservation techniques based on conventional ML algorithms. Then, we have provided the motivations and different robust architectures of FL algorithms for privacy preservation in IoMT networks. Taking it one step further, we have introduced how the performance of FL algorithms can be enhanced using DRL, DNN, and GANs in FL architectures. In the end, we have presented some real-time application and research directions related to FL that can be addressed to further improve the performance of FL from a privacy perspective.



\ifCLASSOPTIONcaptionsoff
  \newpage
\fi

\bibliographystyle{IEEEtran}
\bibliography{bibliography}

\begin{thebibliography}{10}
\providecommand{\url}[1]{#1}
\csname url@samestyle\endcsname
\providecommand{\newblock}{\relax}
\providecommand{\bibinfo}[2]{#2}
\providecommand{\BIBentrySTDinterwordspacing}{\spaceskip=0pt\relax}
\providecommand{\BIBentryALTinterwordstretchfactor}{4}
\providecommand{\BIBentryALTinterwordspacing}{\spaceskip=\fontdimen2\font plus
\BIBentryALTinterwordstretchfactor\fontdimen3\font minus
  \fontdimen4\font\relax}
\providecommand{\BIBforeignlanguage}[2]{{%
\expandafter\ifx\csname l@#1\endcsname\relax
\typeout{** WARNING: IEEEtran.bst: No hyphenation pattern has been}%
\typeout{** loaded for the language `#1'. Using the pattern for}%
\typeout{** the default language instead.}%
\else
\language=\csname l@#1\endcsname
\fi
#2}}
\providecommand{\BIBdecl}{\relax}
\BIBdecl

\bibitem{tai2021trustworthy}
Y.~Tai, B.~Gao, Q.~Li, Z.~Yu, C.~Zhu, and V.~Chang, ``Trustworthy and
  intelligent covid-19 diagnostic iomt through xr and deep-learning-based
  clinic data access,'' \emph{IEEE Internet of Things Journal}, vol.~8, no.~21,
  pp. 15\,965--15\,976, 2021.

\bibitem{mansour2021artificial}
R.~F. Mansour, A.~El~Amraoui, I.~Nouaouri, V.~G. D{\'\i}az, D.~Gupta, and
  S.~Kumar, ``Artificial intelligence and internet of things enabled disease
  diagnosis model for smart healthcare systems,'' \emph{IEEE Access}, vol.~9,
  pp. 45\,137--45\,146, 2021.

\bibitem{naeem2020sdn}
F.~Naeem, M.~Tariq, and H.~V. Poor, ``Sdn-enabled energy-efficient routing
  optimization framework for industrial internet of things,'' \emph{IEEE
  Transactions on Industrial Informatics}, vol.~17, no.~8, pp. 5660--5667,
  2020.

\bibitem{yuan2020federated}
B.~Yuan, S.~Ge, and W.~Xing, ``A federated learning framework for healthcare
  iot devices,'' \emph{arXiv preprint arXiv:2005.05083}, 2020.

\bibitem{sheller2020federated}
M.~J. Sheller, B.~Edwards, G.~A. Reina, J.~Martin, S.~Pati, A.~Kotrotsou,
  M.~Milchenko, W.~Xu, D.~Marcus, R.~R. Colen \emph{et~al.}, ``Federated
  learning in medicine: facilitating multi-institutional collaborations without
  sharing patient data,'' \emph{Scientific reports}, vol.~10, no.~1, pp. 1--12,
  2020.

\bibitem{nguyen2021federated}
D.~C. Nguyen, M.~Ding, P.~N. Pathirana, A.~Seneviratne, J.~Li, D.~Niyato, and
  H.~V. Poor, ``Federated learning for industrial internet of things in future
  industries,'' \emph{IEEE Wireless Communications}, 2021.

\bibitem{khan2021federated}
L.~U. Khan, W.~Saad, Z.~Han, E.~Hossain, and C.~S. Hong, ``Federated learning
  for internet of things: Recent advances, taxonomy, and open challenges,''
  \emph{IEEE Communications Surveys \& Tutorials}, 2021.

\bibitem{xu2021federated}
J.~Xu, B.~S. Glicksberg, C.~Su, P.~Walker, J.~Bian, and F.~Wang, ``Federated
  learning for healthcare informatics,'' \emph{Journal of Healthcare
  Informatics Research}, vol.~5, no.~1, pp. 1--19, 2021.

\bibitem{pham2021fusion}
Q.-V. Pham, K.~Dev, P.~K.~R. Maddikunta, T.~R. Gadekallu, T.~Huynh-The
  \emph{et~al.}, ``Fusion of federated learning and industrial internet of
  things: A survey,'' \emph{arXiv preprint arXiv:2101.00798}, 2021.

\bibitem{rieke2020future}
N.~Rieke, J.~Hancox, W.~Li, F.~Milletari, H.~R. Roth, S.~Albarqouni, S.~Bakas,
  M.~N. Galtier, B.~A. Landman, K.~Maier-Hein \emph{et~al.}, ``The future of
  digital health with federated learning,'' \emph{NPJ digital medicine},
  vol.~3, no.~1, pp. 1--7, 2020.

\bibitem{1}
L.~Tawalbeh, F.~Muheidat, M.~Tawalbeh, M.~Quwaider \emph{et~al.}, ``Iot privacy
  and security: Challenges and solutions,'' \emph{Applied Sciences}, vol.~10,
  no.~12, p. 4102, 2020.

\bibitem{2}
W.~Aman and F.~Kausar, ``Towards a gatewaybased context-aware and self-adaptive
  security management model for iot-based ehealth systems,''
  \emph{International Journal of Advanced Computer Science and Applications},
  vol.~10, no.~1, pp. 280--287, 2019.

\bibitem{3}
M.~M. Ogonji, G.~Okeyo, and J.~M. Wafula, ``A survey on privacy and security of
  internet of things,'' \emph{Computer Science Review}, vol.~38, p. 100312,
  2020.

\bibitem{4}
G.~Hatzivasilis, O.~Soultatos, S.~Ioannidis, C.~Verikoukis, G.~Demetriou, and
  C.~Tsatsoulis, ``Review of security and privacy for the internet of medical
  things (iomt),'' in \emph{2019 15th international conference on distributed
  computing in sensor systems (DCOSS)}.\hskip 1em plus 0.5em minus 0.4em\relax
  IEEE, 2019, pp. 457--464.

\bibitem{5}
F.~Alshehri and G.~Muhammad, ``A comprehensive survey of the internet of things
  (iot) and ai-based smart healthcare.'' \emph{IEEE Access}, vol.~9, no.
  January, pp. 3660--3678, 2021.

\bibitem{6}
R.~Raeside, S.~R. Partridge, A.~Singleton, and J.~Redfern, ``Cardiovascular
  disease prevention in adolescents: ehealth, co-creation, and advocacy,''
  \emph{Medical Sciences}, vol.~7, no.~2, p.~34, 2019.

\bibitem{7}
U.~Ahmad, H.~Song, A.~Bilal, S.~Saleem, and A.~Ullah, ``Securing insulin pump
  system using deep learning and gesture recognition,'' in \emph{2018 17th IEEE
  International Conference On Trust, Security And Privacy In Computing And
  Communications/12th IEEE International Conference On Big Data Science And
  Engineering (TrustCom/BigDataSE)}.\hskip 1em plus 0.5em minus 0.4em\relax
  IEEE, 2018, pp. 1716--1719.

\bibitem{8}
K.~Rannenberg, ``Enabling identity management and privacy in a global context:
  Iso/iec standardization in jtc 1/sc 27/wg 5,'' vol.~11, no.~2, pp. 9--24,
  2018.

\bibitem{9}
A.~M. Conforming, ``Proposal for a privacy impact assessment manual conforming
  to iso/iec 29134: 2017,'' in \emph{Computer Information Systems and
  Industrial Management: 17th International Conference, CISIM 2018, Olomouc,
  Czech Republic, September 27-29, 2018, Proceedings}, vol. 11127.\hskip 1em
  plus 0.5em minus 0.4em\relax Springer, 2018, p. 486.

\bibitem{10}
J.~Brahma and D.~Sadhya, ``Preserving contextual-privacy for smart home iot
  devices with dynamic traffic shaping,'' \emph{IEEE Internet of Things
  Journal}, 2021.

\bibitem{11}
D.~Kerr, K.~Butler-Henderson, and T.~Sahama, ``Security, privacy, and ownership
  issues with the use of wearable health technologies,'' in \emph{Cyber Law,
  Privacy, and Security: Concepts, Methodologies, Tools, and
  Applications}.\hskip 1em plus 0.5em minus 0.4em\relax IGI Global, 2019, pp.
  1629--1644.

\bibitem{12}
C.~Perera, ``Privacy guidelines for internet of things: a cheat sheet,''
  \emph{arXiv preprint arXiv:1708.05261}, 2017.

\bibitem{13}
D.~Chen, H.~Cao, H.~Chen, Z.~Zhu, X.~Qian, W.~Xu, and M.-C. Huang, ``Smart
  insole-based indoor localization system for internet of things
  applications,'' \emph{IEEE Internet of Things Journal}, vol.~6, no.~4, pp.
  7253--7265, 2019.

\bibitem{14}
S.~Zhang, G.~Wang, M.~Z.~A. Bhuiyan, and Q.~Liu, ``A dual privacy preserving
  scheme in continuous location-based services,'' \emph{IEEE Internet of Things
  Journal}, vol.~5, no.~5, pp. 4191--4200, 2018.

\bibitem{15}
T.~Hikita and R.~S. Yamaguchi, ``Preliminary study about advantageous
  trajectory anonymization methods based on population,'' in \emph{2018 10th
  International Conference on Communication Systems \& Networks
  (COMSNETS)}.\hskip 1em plus 0.5em minus 0.4em\relax IEEE, 2018, pp. 492--495.

\bibitem{16}
B.~Niu, X.~Zhu, Q.~Li, J.~Chen, and H.~Li, ``A novel attack to spatial cloaking
  schemes in location-based services,'' \emph{Future Generation Computer
  Systems}, vol.~49, pp. 125--132, 2015.

\bibitem{17}
I.~Ullah and M.~A. Shah, ``A novel model for preserving location privacy in
  internet of things,'' in \emph{2016 22nd International conference on
  automation and computing (ICAC)}.\hskip 1em plus 0.5em minus 0.4em\relax
  IEEE, 2016, pp. 542--547.

\bibitem{18}
R.~Yu, Z.~Bai, L.~Yang, P.~Wang, O.~A. Move, and Y.~Liu, ``A location cloaking
  algorithm based on combinatorial optimization for location-based services in
  5g networks,'' \emph{IEEE Access}, vol.~4, pp. 6515--6527, 2016.

\bibitem{19}
H.~Wen, Y.~Wu, C.~Yang, H.~Duan, and S.~Yu, ``A unified federated learning
  framework for wireless communications: Towards privacy, efficiency, and
  security,'' in \emph{IEEE INFOCOM 2020-IEEE Conference on Computer
  Communications Workshops (INFOCOM WKSHPS)}.\hskip 1em plus 0.5em minus
  0.4em\relax IEEE, 2020, pp. 653--658.

\bibitem{20}
M.~T. Hammi, B.~Hammi, P.~Bellot, and A.~Serhrouchni, ``Bubbles of trust: A
  decentralized blockchain-based authentication system for iot,''
  \emph{Computers \& Security}, vol.~78, pp. 126--142, 2018.

\bibitem{21}
D.~Puthal and S.~P. Mohanty, ``Proof of authentication: Iot-friendly
  blockchains,'' \emph{IEEE Potentials}, vol.~38, no.~1, pp. 26--29, 2018.

\bibitem{22}
M.~A. Ferrag, L.~Maglaras, A.~Ahmim, M.~Derdour, and H.~Janicke, ``Rdtids:
  Rules and decision tree-based intrusion detection system for
  internet-of-things networks,'' \emph{Future internet}, vol.~12, no.~3, p.~44,
  2020.

\bibitem{23}
M.~A. Ferrag, L.~Maglaras, S.~Moschoyiannis, and H.~Janicke, ``Deep learning
  for cyber security intrusion detection: Approaches, datasets, and comparative
  study,'' \emph{Journal of Information Security and Applications}, vol.~50, p.
  102419, 2020.

\bibitem{24}
Y.~Zhang, Q.~Wu, and M.~Shikh-Bahaei, ``Vertical federated learning based
  privacy-preserving cooperative sensing in cognitive radio networks,'' in
  \emph{2020 IEEE Globecom Workshops (GC Wkshps}.\hskip 1em plus 0.5em minus
  0.4em\relax IEEE, 2020, pp. 1--6.

\bibitem{25}
B.~McMahan, E.~Moore, D.~Ramage, S.~Hampson, and B.~A. y~Arcas,
  ``Communication-efficient learning of deep networks from decentralized
  data,'' in \emph{Artificial intelligence and statistics}.\hskip 1em plus
  0.5em minus 0.4em\relax PMLR, 2017, pp. 1273--1282.

\bibitem{26}
L.~Feng, Y.~Zhao, S.~Guo, X.~Qiu, W.~Li, and P.~Yu, ``Blockchain-based
  asynchronous federated learning for internet of things,'' \emph{IEEE
  Transactions on Computers}, 2021.

\bibitem{27}
Z.~Xiong, Z.~Cai, D.~Takabi, and W.~Li, ``Privacy threat and defense for
  federated learning with non-iid data in aiot,'' \emph{IEEE Transactions on
  Industrial Informatics}, 2021.

\bibitem{28}
M.~Ali, H.~Karimipour, and M.~Tariq, ``Integration of blockchain and federated
  learning for internet of things: Recent advances and future challenges,''
  \emph{Computers \& Security}, p. 102355, 2021.

\bibitem{29}
Z.~Yang, M.~Chen, W.~Saad, C.~S. Hong, and M.~Shikh-Bahaei, ``Energy efficient
  federated learning over wireless communication networks,'' \emph{IEEE
  Transactions on Wireless Communications}, vol.~20, no.~3, pp. 1935--1949,
  2020.

\bibitem{30}
Y.~Lu, X.~Huang, K.~Zhang, S.~Maharjan, and Y.~Zhang, ``Low-latency federated
  learning and blockchain for edge association in digital twin empowered 6g
  networks,'' \emph{IEEE Transactions on Industrial Informatics}, vol.~17,
  no.~7, pp. 5098--5107, 2020.

\bibitem{31}
A.~Hard, K.~Rao, R.~Mathews, S.~Ramaswamy, F.~Beaufays, S.~Augenstein,
  H.~Eichner, C.~Kiddon, and D.~Ramage, ``Federated learning for mobile
  keyboard prediction,'' \emph{arXiv preprint arXiv:1811.03604}, 2018.

\bibitem{32}
R.~Kumar, A.~A. Khan, J.~Kumar, A.~Zakria, N.~A. Golilarz, S.~Zhang, Y.~Ting,
  C.~Zheng, and W.~Wang, ``Blockchain-federated-learning and deep learning
  models for covid-19 detection using ct imaging,'' \emph{IEEE Sensors
  Journal}, 2021.

\bibitem{33}
M.~H. ur~Rehman, A.~M. Dirir, K.~Salah, E.~Damiani, and D.~S. Center,
  ``Trustfed: A framework for fair and trustworthy cross-device federated
  learning in iiot,'' \emph{IEEE Transactions on Industrial Informatics}, 2021.

\bibitem{34}
M.~A. Ferrag, O.~Friha, L.~Maglaras, H.~Janicke, and L.~Shu, ``Federated deep
  learning for cyber security in the internet of things: Concepts,
  applications, and experimental analysis,'' \emph{IEEE Access}, vol.~9, pp.
  138\,509--138\,542, 2021.

\bibitem{35}
D.~C. Nguyen, Q.-V. Pham, P.~N. Pathirana, M.~Ding, A.~Seneviratne, Z.~Lin,
  O.~A. Dobre, and W.-J. Hwang, ``Federated learning for smart healthcare: A
  survey,'' \emph{arXiv preprint arXiv:2111.08834}, 2021.

\bibitem{36}
C.~Xu, N.~Wang, L.~Zhu, K.~Sharif, and C.~Zhang, ``Achieving searchable and
  privacy-preserving data sharing for cloud-assisted e-healthcare system,''
  \emph{IEEE Internet of Things Journal}, vol.~6, no.~5, pp. 8345--8356, 2019.

\bibitem{37}
M.~Staffa, L.~Sgaglione, G.~Mazzeo, L.~Coppolino, S.~D'Antonio, L.~Romano,
  E.~Gelenbe, O.~Stan, S.~Carpov, E.~Grivas \emph{et~al.}, ``An openncp-based
  solution for secure ehealth data exchange,'' \emph{Journal of Network and
  Computer Applications}, vol. 116, pp. 65--85, 2018.

\bibitem{38}
L.~U. Khan, S.~R. Pandey, N.~H. Tran, W.~Saad, Z.~Han, M.~N. Nguyen, and C.~S.
  Hong, ``Federated learning for edge networks: Resource optimization and
  incentive mechanism,'' \emph{IEEE Communications Magazine}, vol.~58, no.~10,
  pp. 88--93, 2020.

\bibitem{39}
L.~Zhu and S.~Han, ``Deep leakage from gradients,'' in \emph{Federated
  learning}.\hskip 1em plus 0.5em minus 0.4em\relax Springer, 2020, pp. 17--31.

\bibitem{40}
M.~Nasr, R.~Shokri, and A.~Houmansadr, ``Machine learning with membership
  privacy using adversarial regularization,'' in \emph{Proceedings of the 2018
  ACM SIGSAC Conference on Computer and Communications Security}, 2018, pp.
  634--646.

\bibitem{agrawal2021federated}
S.~Agrawal, S.~Sarkar, O.~Aouedi, G.~Yenduri, K.~Piamrat, S.~Bhattacharya,
  P.~K.~R. Maddikunta, and T.~R. Gadekallu, ``Federated learning for intrusion
  detection system: Concepts, challenges and future directions,'' \emph{arXiv
  preprint arXiv:2106.09527}, 2021.

\bibitem{tahir2021experience}
B.~Tahir, A.~Jolfaei, and M.~Tariq, ``Experience driven attack design and
  federated learning based intrusion detection in industry 4.0,'' \emph{IEEE
  Transactions on Industrial Informatics}, 2021.

\bibitem{41}
B.~Hitaj, G.~Ateniese, and F.~Perez-Cruz, ``Deep models under the gan:
  information leakage from collaborative deep learning,'' in \emph{Proceedings
  of the 2017 ACM SIGSAC Conference on Computer and Communications Security},
  2017, pp. 603--618.

\bibitem{42}
Y.~Dong, X.~Chen, L.~Shen, and D.~Wang, ``Eastfly: Efficient and secure ternary
  federated learning,'' \emph{Computers \& Security}, vol.~94, p. 101824, 2020.

\bibitem{43}
J.~Tan, Y.-C. Liang, N.~C. Luong, and D.~Niyato, ``Toward smart security
  enhancement of federated learning networks,'' \emph{IEEE Network}, vol.~35,
  no.~1, pp. 340--347, 2020.

\bibitem{44}
Y.~Zhao, J.~Chen, J.~Zhang, D.~Wu, M.~Blumenstein, and S.~Yu, ``Detecting and
  mitigating poisoning attacks in federated learning using generative
  adversarial networks,'' \emph{Concurrency and Computation: Practice and
  Experience}, p. e5906, 2020.

\bibitem{45}
Q.~Wang, Y.~Guo, X.~Wang, T.~Ji, L.~Yu, and P.~Li, ``Ai at the edge:
  Blockchain-empowered secure multiparty learning with heterogeneous models,''
  \emph{IEEE Internet of Things Journal}, vol.~7, no.~10, pp. 9600--9610, 2020.

\bibitem{46}
W.~Sun, S.~Lei, L.~Wang, Z.~Liu, and Y.~Zhang, ``Adaptive federated learning
  and digital twin for industrial internet of things,'' \emph{IEEE Transactions
  on Industrial Informatics}, vol.~17, no.~8, pp. 5605--5614, 2020.

\bibitem{47}
Y.~Wang, Z.~Su, N.~Zhang, and A.~Benslimane, ``Learning in the air: Secure
  federated learning for uav-assisted crowdsensing,'' \emph{IEEE Transactions
  on network science and engineering}, 2020.

\bibitem{48}
M.~Hao, H.~Li, X.~Luo, G.~Xu, H.~Yang, and S.~Liu, ``Efficient and
  privacy-enhanced federated learning for industrial artificial intelligence,''
  \emph{IEEE Transactions on Industrial Informatics}, vol.~16, no.~10, pp.
  6532--6542, 2019.

\bibitem{49}
Y.~Liu, Z.~Ma, X.~Liu, S.~Ma, S.~Nepal, R.~H. Deng, and K.~Ren, ``Boosting
  privately: Federated extreme gradient boosting for mobile crowdsensing,'' in
  \emph{2020 IEEE 40th International Conference on Distributed Computing
  Systems (ICDCS)}.\hskip 1em plus 0.5em minus 0.4em\relax IEEE, 2020, pp.
  1--11.

\bibitem{rm2020effective}
S.~P. RM, P.~K.~R. Maddikunta, M.~Parimala, S.~Koppu, T.~R. Gadekallu, C.~L.
  Chowdhary, and M.~Alazab, ``An effective feature engineering for dnn using
  hybrid pca-gwo for intrusion detection in iomt architecture,'' \emph{Computer
  Communications}, vol. 160, pp. 139--149, 2020.

\bibitem{50}
M.~Wu, D.~Ye, J.~Ding, Y.~Guo, R.~Yu, and M.~Pan, ``Incentivizing
  differentially private federated learning: A multidimensional contract
  approach,'' \emph{IEEE Internet of Things Journal}, vol.~8, no.~13, pp.
  10\,639--10\,651, 2021.

\bibitem{51}
J.~Zhao, X.~Zhu, J.~Wang, and J.~Xiao, ``Efficient client contribution
  evaluation for horizontal federated learning,'' in \emph{ICASSP 2021-2021
  IEEE International Conference on Acoustics, Speech and Signal Processing
  (ICASSP)}.\hskip 1em plus 0.5em minus 0.4em\relax IEEE, 2021, pp. 3060--3064.

\bibitem{52}
O.~Choudhury, A.~Gkoulalas-Divanis, T.~Salonidis, I.~Sylla, Y.~Park, G.~Hsu,
  and A.~Das, ``Differential privacy-enabled federated learning for sensitive
  health data,'' \emph{arXiv preprint arXiv:1910.02578}, 2019.

\bibitem{53}
Y.~Zhan, J.~Zhang, Z.~Hong, L.~Wu, P.~Li, and S.~Guo, ``A survey of incentive
  mechanism design for federated learning,'' \emph{IEEE Transactions on
  Emerging Topics in Computing}, 2021.

\bibitem{54}
Y.~Sarikaya and O.~Ercetin, ``Motivating workers in federated learning: A
  stackelberg game perspective,'' \emph{IEEE Networking Letters}, vol.~2,
  no.~1, pp. 23--27, 2019.

\bibitem{55}
Y.~Zhan, P.~Li, Z.~Qu, D.~Zeng, and S.~Guo, ``A learning-based incentive
  mechanism for federated learning,'' \emph{IEEE Internet of Things Journal},
  vol.~7, no.~7, pp. 6360--6368, 2020.

\bibitem{56}
M.~Tang and V.~W. Wong, ``An incentive mechanism for cross-silo federated
  learning: A public goods perspective,'' in \emph{IEEE INFOCOM 2021-IEEE
  Conference on Computer Communications}.\hskip 1em plus 0.5em minus
  0.4em\relax IEEE, 2021, pp. 1--10.

\bibitem{57}
J.~Zhao, X.~Zhu, J.~Wang, and J.~Xiao, ``Efficient client contribution
  evaluation for horizontal federated learning,'' in \emph{ICASSP 2021-2021
  IEEE International Conference on Acoustics, Speech and Signal Processing
  (ICASSP)}.\hskip 1em plus 0.5em minus 0.4em\relax IEEE, 2021, pp. 3060--3064.

\bibitem{58}
D.~Gupta, O.~Kayode, S.~Bhatt, M.~Gupta, and A.~S. Tosun, ``Hierarchical
  federated learning based anomaly detection using digital twins for smart
  healthcare,'' \emph{arXiv preprint arXiv:2111.12241}, 2021.

\bibitem{59}
M.~Hao, H.~Li, G.~Xu, Z.~Liu, and Z.~Chen, ``Privacy-aware and resource-saving
  collaborative learning for healthcare in cloud computing,'' in \emph{ICC
  2020-2020 IEEE International Conference on Communications (ICC)}.\hskip 1em
  plus 0.5em minus 0.4em\relax IEEE, 2020, pp. 1--6.

\bibitem{60}
Z.~Yan, J.~Wicaksana, Z.~Wang, X.~Yang, and K.-T. Cheng, ``Variation-aware
  federated learning with multi-source decentralized medical image data,''
  \emph{IEEE Journal of Biomedical and Health Informatics}, vol.~25, no.~7, pp.
  2615--2628, 2020.

\bibitem{61}
W.~Li, F.~Milletar{\`\i}, D.~Xu, N.~Rieke, J.~Hancox, W.~Zhu, M.~Baust,
  Y.~Cheng, S.~Ourselin, M.~J. Cardoso \emph{et~al.}, ``Privacy-preserving
  federated brain tumour segmentation,'' in \emph{International workshop on
  machine learning in medical imaging}.\hskip 1em plus 0.5em minus 0.4em\relax
  Springer, 2019, pp. 133--141.

\bibitem{62}
D.~C. Nguyen, M.~Ding, P.~N. Pathirana, and A.~Seneviratne, ``Blockchain and
  ai-based solutions to combat coronavirus (covid-19)-like epidemics: A
  survey,'' \emph{Ieee Access}, vol.~9, pp. 95\,730--95\,753, 2021.

\bibitem{63}
Q.-V. Pham, D.~C. Nguyen, T.~Huynh-The, W.-J. Hwang, and P.~N. Pathirana,
  ``Artificial intelligence (ai) and big data for coronavirus (covid-19)
  pandemic: A survey on the state-of-the-arts,'' \emph{IEEE access}, vol.~8, p.
  130820, 2020.

\bibitem{64}
M.~Loey, F.~Smarandache, and N.~E. M~Khalifa, ``Within the lack of chest
  covid-19 x-ray dataset: a novel detection model based on gan and deep
  transfer learning,'' \emph{Symmetry}, vol.~12, no.~4, p. 651, 2020.

\bibitem{65}
R.~Kumar, A.~A. Khan, J.~Kumar, N.~A. Golilarz, S.~Zhang, Y.~Ting, C.~Zheng,
  W.~Wang \emph{et~al.}, ``Blockchain-federated-learning and deep learning
  models for covid-19 detection using ct imaging,'' \emph{IEEE Sensors
  Journal}, vol.~21, no.~14, pp. 16\,301--16\,314, 2021.

\bibitem{66}
W.~Zhang, T.~Zhou, Q.~Lu, X.~Wang, C.~Zhu, H.~Sun, Z.~Wang, S.~K. Lo, and F.-Y.
  Wang, ``Dynamic-fusion-based federated learning for covid-19 detection,''
  \emph{IEEE Internet of Things Journal}, vol.~8, no.~21, pp. 15\,884--15\,891,
  2021.

\bibitem{67}
F.~Qiang, T.~Lixin, L.~Richard \emph{et~al.}, ``White paper-ieee federated
  machine learning,'' 2021.

\bibitem{68}
C.~A. Choquette-Choo, N.~Dullerud, A.~Dziedzic, Y.~Zhang, S.~Jha, N.~Papernot,
  and X.~Wang, ``Capc learning: Confidential and private collaborative
  learning,'' \emph{arXiv preprint arXiv:2102.05188}, 2021.

\bibitem{69}
C.~De~Alwis, A.~Kalla, Q.-V. Pham, P.~Kumar, K.~Dev, W.-J. Hwang, and
  M.~Liyanage, ``Survey on 6g frontiers: Trends, applications, requirements,
  technologies and future research,'' \emph{IEEE Open Journal of the
  Communications Society}, vol.~2, pp. 836--886, 2021.

\end{thebibliography}

\end{document}